\documentclass[twopage,12pt]{article}

\newcommand{\vekk}[1]{}
\usepackage[utf8]{inputenc}

\input{def.tex}

\begin{document}

\title{Discussion of
`Nonparametric generalized fiducial inference for survival
functions under censoring'}

\author{G. Taraldsen and B.H. Lindqvist \\
\\
Department of Mathematical Sciences \\
Norwegian University of Science and Technology\\
NTNU, NO-7491 Trondheim, Norway\\
\\
Gunnar.Taraldsen@ntnu.no and  Bo.Lindqvist@ntnu.no}

\maketitle

\begin{abstract}
The following discussion is inspired by the paper
{\em Nonparametric generalized fiducial inference for survival
functions under censoring} by Cui and Hannig.
The discussion consists of comments on the results,
but also indicates it's importance more generally in the context of fiducial inference. 
A two page introduction to fiducial inference is given to provide a context.\\

\end{abstract}
  
\section{Fiducial inference}

We expect that many readers are not familiar with fiducial inference.
This is in contrast to the well founded alternatives given by Bayesian and classical inference
known to every statistician today.
Fiducial inference has not yet been established as a general theory,
but there has been considerable progress on this during the last decades,
as also demonstrated by \citet{CuiHannig19nonparFiducial}.
To discuss their contribution we need to provide a context given
by fiducial inference as we see it today.

The original fiducial argument of \citet[p.532]{Fisher30} starts by considering the relation
\be{1}
u = F(x)
\ee
where $F$ is the cumulative distribution function for the observation $x$.
Fisher considers in particular the case where $x$ is the empirical correlation
of a sample of size $n$ from the bivariate Gaussian distribution.
In this case $F$ is strictly decreasing from $1$ down to $0$ as a function
of the unknown % parameter given by the
correlation $\theta$.
From this,
Fisher argues that $1 - F(x \st \theta)$ is the cumulative
fiducial distribution for $\theta$,
and that $\pi^x (\theta) = -\partial_\theta F (x \st \theta)$
is the fiducial density of $\theta$ given $x$.

Fisher's argument uses the fact that equation~(\ref{eq1}) gives a correspondence
between a uniform law for $u$ and the sampling law for $x$.
The argument explains, in fact, that the percentiles of the fiducial distribution give
confidence intervals, and hence that the fiducial distribution is a confidence distribution
in this case.
Even though Fisher himself abandoned this interpretation in later works,
it must be seen as
one of the pioneering works that lead to the theory of confidence intervals and
hypothesis testing as used today.
It is, as far as we know, the first paper that calculates
exact confidence intervals and explain them as such.
% This, and slight extensions of Fisher's original argument, are elaborated on by
% \citet[p.141-144]{TaraldsenLindqvist18fiducial}.

Fiducial inference, in the version considered here,
is given by replacing the relation~(\ref{eq1}) by a {\em fiducial model}
\be{2}
x = \theta u
\ee
%
%The right-hand side is simply a very economic notation for a general function of the
%two variables $\theta$ and $u$.
This economic notation is used by \citet[p.1055]{DawidStone82}
when they define a {\em functional model}.
It is a generalization of the {\em structural models} of \citet{FRASER} who considers the
case where the model space $\Omega_\Theta$ is a group,
and $\theta u$ is the action of $\theta$ on $u$.
\citet[eq.1]{CuiHannig19nonparFiducial} refer to equation~(\ref{eq2}) as a
%\citet{CuiHannig19nonparFiducial} refers to equation~(\ref{eq2}) as a
{\em data generating equation}.
Samples from a known distribution for $u$ gives samples from the distribution
of the observation $x$.
In modern statistics,
the possibility of simulating data from a statistical model
is most central, and any such algorithm is in fact a fiducial model.
% Strictly speaking, a fiducial model is then defined by equation~(\ref{eq2}) and
% a specification of a distribution for $u$. % \citep{TaraldsenLindqvist13fidopt}.

Equation~(\ref{eq1}) can be inverted to give $x = \theta u = F^{-1} (u)$,
where $F$ depends on $\theta$.
Fisher's initial model is hence a special case of a fiducial model.
Consider for a moment the following problem:
\begin{quote}
  {\it
The observation $x$ is given and known to be generated
from the fiducial model~(\ref{eq2})
by sampling $u$ from a known distribution.
How would You quantify Your uncertainty about the unknown model parameter $\theta$?
}
\end{quote}
It is clear that both $u$ and $\theta$ are still uncertain,
and it is reasonable, we claim, to quantify these uncertainties by
a joint distribution for $(u,\theta)$ such that equation~(\ref{eq2}) holds.
Define $\theta = x u^{-1}$ to be a measurable selection solution of
equation~(\ref{eq2}) for those $(x,u)$ that allows a solution.
Assume, as we will exemplify below, that there exists a fiducial distribution for
$u^x$ derived from the original distribution of $u$ and the observation $x$.
A fiducial distribution for the model $\theta$ can then be defined to be 
the distribution of % $\theta^x = x [u^x]^{-1}$.
%\citep{TaraldsenLindqvist13fidopt}.
\be{3}
\theta^x = x \left(u^x \right)^{-1}
\ee
The fiducial distribution quantifies the uncertainty of $\theta$ given the assumed
fiducial model and given the observation $x$.
This interpretation of the fiducial %,
% as discussed in more detail by \citet[p.141-144]{TaraldsenLindqvist18fiducial},
is what \citet[p.54-55]{FISHER} aimed at in his final writing on this:
\begin{quote}
{\it
  By contrast, the fiducial argument uses the observations only to change the
  logical status of the parameter from one in
  which nothing is known of it, and no probability statement about it can be made,
  to the status of a random variable having a well-defined distribution.}
\end{quote}
The correlation coefficient example treated initially by Fisher is such that
the fiducial equation~(\ref{eq2}) defines a one-one correspondence between any two
variables when the third is fixed.
In this case, a simple fiducial model, 
the distribution of $u^x$ can be set equal to the original distribution of $u$.
Fiducial samples are obtained simply by solving the fiducial equation for each sample $u$
and returning the solution $\theta^x = x u^{-1}$.

Another example is given by $x = \theta u = \theta + u$,
where $\theta$ is an element of a subspace $\Omega_\Theta$ of a Hilbert space $\Omega_X$.
An important class of problems is obtained by letting
$\Omega_\Theta$ be the image space of the design matrix in linear regression.
In this case, the fiducial equation will fail to have solutions for all $(x,u)$.
Let $P$ be the orthogonal projection on $\Omega_\Theta$,
and let $Q = 1 - P$.
Define the law of $u^x$ to be the conditional law of $u$
given $Q u = Q x$.
The fiducial is then
$\theta^x = x [u^x]^{-1} = x - u^x$.

The previous example includes the general case of a location parameter,
and in particular inference based on sampling from the
Gaussian distribution with unknown mean and known variance.
As demonstrated by \citet{FRASER},
this can be seen as a particular
case of a group $\Omega_\Theta$ acting on the observation space $\Omega_X$,
and cases with unknown variance can also be included by considering other group actions.
It follows in these cases, as also for the simple fiducial models,
that the fiducial is a confidence distribution.
Furthermore, \citet{TaraldsenLindqvist13fidopt} have proved that
classical optimal actions, if they exist, are determined by the
fiducial if the loss is invariant.
Incidentally, the previous also exemplify a nonparametric fiducial in the
sense given by an infinite dimensional $\Omega_\Theta$.

The previous indicate that a fiducial model~(\ref{eq2}) can be used
to obtain a distribution with interpretation similar to a Bayesian posterior
as intended originally by Fisher.
It also show that confidence distributions and classical optimal
actions can be obtained by fiducial arguments.
Finally,
a  fiducial model~(\ref{eq2}) can also be used
as a method for sampling from a Bayesian posterior.
In a Bayesian set-up the joint distribution of $(u,\theta)$
is specified,
and the distribution of $u$ used above must be identified
with the conditional distribution of $u$ given $\theta$.
% This includes cases with an improper prior for $\theta$
% \citep{TaraldsenTuftoLindqvist18improper}.
Sampling from the posterior can be done by sampling
$u$ conditionally given $x$ and then $\theta$ given $(u,x)$.
In the case of group actions with prior equal to the right invariant prior
this gives that the posterior coincides with the fiducial.
% This result holds more generally if the prior law for $\theta$ is such that
% the law of $\theta u$ does not depend on the fixed value of $u$
% \citep{TaraldsenLindqvist15fidpost}.

\section{The results in the paper and future research}

\citet{CuiHannig19nonparFiducial} consider 
failure distributions based on right censored data in a nonparametric case.
For simplicity, and since we will focus on the theoretical principles,
we will focus on the uncensored case.
Before leaving the censored case we will emphasize its importance in applications,
and add, as we see it, that the fiducial model for this case
is most natural.
The ease of including this in the analysis is by itself a
most convincing argument for the success of fiducial inference as
demonstrated by \citet{CuiHannig19nonparFiducial}.

The obvious choice, in retrospect, is to base nonparametric fiducial inference
on Fisher's original fiducial relation in equation~(\ref{eq1}).
The data is given by an ordered sample $x$ that obeys the
fiducial relation $u_{i} = F(x_{i})$,
or equivalently the fiducial model $x_i = F^{-1} (u_i)$.
Here
$u_1 \le \cdots \le u_n$ is the order statistic 
of a random sample from the uniform distribution on $[0,1]$.
A fiducial distribution for the unknown cumulative distribution function $F$
is given by a measurable selection solution of this fiducial relation.
We can and will restrict attention to the case where it is assumed that
$F$ is absolutely continuous in accordance with 
\citet[Assumption 2]{CuiHannig19nonparFiducial}.
In this case it follows hence that the fiducial distribution
for $u^x$ equals the original distribution for $u$ as
in Fishers original fiducial argument for the correlation coefficient.
In contrast to Fishers original argument there is here an infinity of
possible randomized measurable selection solutions.
It can, additionally,
be observed that the given fiducial model is equivalent with
a group model $x = \theta v$:
$\; \Omega_\Theta$ is the group of increasing and differentiable transformations $\theta$
of the positive real line and
$v_1 \le \cdots \le v_n$ is the order statistic 
of a random sample from the standard exponential distribution.
%This is a measure given by $F$ which is jointly measurable as a function of
% the Monte Carlo variable $u$ and the data $x$.

A particular absolutely continuous fiducial $F^I$ is determined by 
log-linear interpolation as described by
\citet{CuiHannig19nonparFiducial}.
This gives fiducial distributions for any parameters of interest,
and in particular for $F(x)$ for a fixed $x$ and the
percentiles $x_\alpha$ for a fixed $\alpha$.
The case with $k$ samples can be
treated similarly by the joint fiducial for $F_1, \ldots, F_k$.
It is straightforward, in principle, to calculate corresponding fiducial intervals or regions and
corresponding fiducial $p$-values.
This is exemplified by \citet{CuiHannig19nonparFiducial} by a series of examples for $k=1,2$,
and good frequentist properties are demonstrated as compared with existing methodology.
The group model structure opens the question:
Is optimal equivariant inference possible?

The demonstrations, and the previous two paragraphs,
constitute, in our opinion, the main message of the paper.
Many more examples can, and should, be published based on concrete applied problems,
and the indicated natural route for nonparametric inference.
An alternative approach is to take your favorite book on nonparametric inference and
implement and experiment with corresponding fiducial solutions.
% The remainder of our discussion will be on more technical aspects of the analysis.

Proofs of stated coverage in the finite sample case are absent,
but for $k > 1$ this can be expected to be a long standing challenge as even the
Behrens-Fisher problem remains unsettled.
The $k=1$ case seems possible to analyse completely, and the methodology should then be
compared with similar results for the uncensored case presented by
\citet[Chap.11]{SchwederHjort16book}.
It should be noted that \citet{SchwederHjort16book}
only consider confidence distributions for real valued parameters,
and not for the unknown $F$ itself.
It is, in fact, unknown if the fiducial for $F$ is a confidence distribution in a strict sense.
The group model structure gives a starting point for investigating this further. 
% A related fundamental question is:
% Does a prior for $F$ exist such that
% the resulting posterior is fiducial?

All of these questions are related to the choice of a measurable selection solution.
Is there a natural choice? Is there a best choice?
This question should be investigated in concrete data situations.
It can be observed that the choice $F^I$ is quick and convenient,
but each realization is so special that it is not realistic in most situations.
An alternative, which is still quick and convenient, is given by
monotonic spline interpolation.
The fiducial distribution given by $F^I$ has defects when considered as
a fiducial distribution for $F$,
but the simulations demonstrate that resulting finite dimensional fiducials
of certain focus parameters have excellent properties.

In summary, what is the possible role of the fiducial
argument and distribution?
The following Bayesian-Fiducial-Frequentist list give guidance:
%
%\begin{center}
%\hspace{10ex}\parbox{\textwidth}{
%\begin{minipage}[][][c]{\textwidth}
\begin{description}
\item[(B)] Alternative algorithms for Bayesian analysis.
\item[(F)] A posterior fiducial state interpreted as Fisher intended.
\item[(F)] Alternative algorithms for frequentist analysis.
\end{description}
%\end{minipage}
% }
%\end{center}
% 
All of this, seen in retrospect,
is excellently presented and exemplified by \citet{FRASER} for classical linear models.
We believe that \citet{CuiHannig19nonparFiducial} have taken the first important step
for similar results in the nonparametric case.
Their main technical result proves that the nonparametric fiducial
is asymptotically a confidence distribution.

\section{Conclusion}

We take the opportunity of  expressing our thanks for the
invitation to comment on the interesting and thought-provoking paper
by \citet{CuiHannig19nonparFiducial}.
This paper will serve as motivation for further developments of
the theory of fiducial inference as initiated by Fisher in the
{\em Inverse probability} paper from 1930.
The importance of the 1930 paper by Fisher,
lies, according to
\citet{Fisher50contributions},
in retrospect,
in setting forth a new mode of reasoning from observations to their hypothetical causes.
We congratulate Cui and Hannig with a successful demonstration of
a fiducial argument in a nonparametric problem.
In conclusion, we can wholeheartedly and repeatedly agree with
\citet[p.107]{Efron98}:
\begin{quote}
{\it
This  is all  quite  speculative,  but  here  is  a safe
prediction for  the  21st  century:  statisticians  will  be
asked  to solve bigger  and  more complicated problems.
I believe  that  there  is  a good chance  that
objective Bayes  methods  will be developed for  such
problems,  and  that  something  like  fiducial
inference will  play  an  important  role  in  this
development.  Maybe  Fisher's  biggest  blunder  will become a
big  hit  in  the  21st  century!}
\end{quote}
Additionally,
we believe that the
addition of nonparametric fiducial inference,
as introduced by \citet{CuiHannig19nonparFiducial},
will play an important part of this adventure.

\bibliographystyle{chicago}
\bibliography{bib,gtaralds}

\end{document}